\documentclass[12pt,final]{iopart}
\usepackage{iopams}
\usepackage{graphicx}
\begin{document}

\title[Linear and Nonlinear Rogue Wave Statistics in the Presence of Random Currents]{Linear and Nonlinear Rogue Wave Statistics in the Presence of Random Currents}

\author{L. H. Ying,$^1$ Z. Zhuang$^2$, E. J. Heller,$^3$ and L. Kaplan,$^1$}

\address{$^1$Department of Physics and Engineering Physics, Tulane University, New Orleans, Louisiana 70118, USA}

\address{$^2$School of Electrical and Computer Engineering, Cornell University, Ithaca, New York 14853, USA}

\address{$^3$Department of Physics and Department of Chemistry and Chemical Biology, Harvard University,
Cambridge, Massachusetts 02138, USA}

\begin{abstract}
We review recent progress in modeling the probability distribution of wave heights in the deep ocean as a function of a small number of parameters describing the local sea state. Both linear and nonlinear mechanisms of rogue wave formation are considered. First, we show that when the average wave steepness is small and nonlinear wave effects are subleading, the wave height distribution is well explained by a single ``freak index'' parameter, which describes the strength of (linear) wave scattering by random currents relative to the angular spread of the incoming random sea. When the average steepness is large, the wave height distribution takes a very similar functional form, but the key variables determining the probability distribution are the steepness, and the angular and frequency spread of the incoming waves. Finally, even greater probability of extreme wave formation is predicted when linear and nonlinear effects are acting together. 
\end{abstract}

\maketitle

\section{Introduction}
\label{secintro}

Tales of freak waves by lucky survivors used to be taken with a large grain of salt. Were the sailors making excuses for bad seamanship? The first such wave to be measured directly was the famous New Year's wave in 1995~\cite{trulsen97}. With modern cameras and video, not to mention satellites~\cite{dankert, schulz04}, it is no longer controversial that freak or rogue extreme waves exist on the world's great oceans~\cite{dysthe08,mallory,kharif}.

Any realistic seaway (an irregular, moderate to rough sea) is comprised of a superposition of waves differing in wavelength and direction, with random relative phases. Supposing that the dispersion in wavelengths is not too large, and assuming uniform sampling, Longuet-Higgins~\cite{lh} exploited the central limit theorem to derive a large number of statistical properties of such wave superpositions, including of course wave height distributions. From this viewpoint, extreme waves are the result of unlucky coherent addition of plane waves corresponding to the tail of the Gaussian distribution (see Eq.~(\ref{prayleigh}) below). As explained below, wave heights greater than about $4 \,\sigma$ in the tail of the Gaussian are classified as extreme. The problem has become how to explain why the observed number of rogue wave events is greater than the number $4\,\sigma$ out in the Longuet-Higgins theory.
 
For the following discussion, it is important to understand why a 20 meter wave in a sea where the significant wave height (SWH, defined as the average height of the highest one third of the waves) is 18 meters is far less onerous than a 20 meter wave where the SWH is 8 meters. An established seaway of uniform energy density (uniform if averaged over an area large compared to the typical wavelength) is ``accommodated'' over time and distance, through nonlinear energy transfer mechanisms. Seaways of higher energy density develop correspondingly longer wave periods and wavelengths, even with no further wind forcing, keeping wave steepness under control as a result.

This accommodation process is one of the ways nonlinear processes are implicitly lurking behind ``linear'' theories, in that the input into the linear theories, i.e., the SWH, the period, dispersion in direction, and dispersion in wavelength are all the result of prior nonlinear processes. A 20 meter wave in a sea of SWH 8 meters is necessarily very steep, possibly breaking, with a deep narrow trough preceding it. The tendency for steep waves to break is an often devastating blow just as the ship is sailing over an unusually deep trough before meeting the crest. 

Observational evidence has shown that the linear Longuet-Higgins theory is too simplistic~\cite{dysthe08}. Recent advances in technology have allowed multiple wave tank experiments and field observations to be conducted, confirming the need for a more realistic theory to explain the results~\cite{forristall00, onorato04}. An obvious correction is to incorporate nonlinear wave evolution at every stage, rather than split the process into an implicit nonlinear preparation of the seaway followed by linear propagation. Clearly the exact evolution is always nonlinear to some extent, but the key is to introduce nonlinearities at the right moment and in an insightful and computable way. Realistic fully nonlinear computations wave by wave over large areas are very challenging, but initial attempts have been made to simulate the ocean surface using the full Euler equation both on large scales~\cite{tanaka01} and over smaller areas~\cite{gibson05, gibson07}. 

Surprisingly, investigation of nonlinear effects is actually not the next logical step needed to improve upon the Longuet-Higgins model. Indeed the full {\em linear} statistical theory had not been given, for the reason that uniform sampling assumed by Longuet-Higgins is not justified. A nonuniform sampling theory, which does not assume that the energy density is uniformly distributed over all space, is possible and is still ``linear.'' Moreover, the parameters governing a nonuniform sampling are knowable. Inspired by the work of White and Fornberg~\cite{wf}, the present authors showed that current eddies commonly present in the oceans are sufficient to cause the time-averaged wave intensity to be spatially non-uniform, and to exhibit ``hot spots'' and ``cold spots'' some tens to hundreds of kilometers down flow from the eddies. We emphasize that in terms of wave evolution, the refraction leading to the patchy energy density is purely linear evolution. The key ideas are (1) that waves suddenly entering a high energy patch are not accommodated to it and grow steep and dangerous, and (2) the process is still probabilistic and the central limit theorem still applies, with the appropriate sampling over a nonuniform distribution. The high-energy patches will skew the tails of the wave height distribution, perhaps by orders of magnitude. This was the main point in reference~\cite{hkd}.

There is no denying the importance of nonlinear effects in wave evolution, and a full theory should certainly include them. On the other hand a nonlinear theory that fails to account for patchy energy density is missing an important, even crucial effect. The linear theory needs to be supplemented by nonlinear effects however, since the accommodation of the waves to the presence of patchy energy density needs to be considered. It is our goal here to review progress along these lines and point the way to a more complete theory. We first review the nonuniform sampling linear theory and then discuss newer simulations using the nonlinear Schr\"odinger equation (NLSE). Finally, we show that even larger rogue wave formation probabilities are predicted when linear and nonlinear formation mechanisms are acting in concert.

Rogue wave modeling would benefit greatly from a comprehensive, accurate, and unbiased global record of extreme wave events, supplemented by data on local ocean conditions, including current strength, SWH, steepness, and the angular and spectral spread of the sea state. Such a record, not available at present, would allow for direct statistical tests of linear and nonlinear theories of rogue wave formation. Anecdotal evidence does suggest that rogue waves may be especially prevalent in regions of strong current, including the Gulf Stream, the Kuroshio Current, and especially the Agulhas Current off the coast of South Africa. Consequently, the Agulhas Current in particular has attracted much attention in rogue wave research~\cite{mallory,agulhas}. However, anecdotal evidence from ships and even oil platform measurements cannot provide a systematic, unbiased, and statistically valid record that would support a correlation between possibly relevant variables and rogue wave formation probability. Instead, satellite-based synthetic aperture radar (SAR)~\cite{dankert, schulz04} is currently the only method that shows potential for monitoring the ocean globally with single-wave resolution, but validating the surface elevations obtained by SAR is a challenge. The SAR imaging mechanism is nonlinear, and may yield a distorted image of the ocean wave field; the nonlinearity is of course of particular concern for extreme events~\cite{ja}. Recently, an empirical approach has been proposed that may accurately obtain parameters such as the SWH from SAR data~\cite{sskl}.

\section{Linear Wave Model in Presence of Currents}
\label{seclinear}

\subsection{Ray Density Statistics}
\label{secrayinten}

To understand the physics of linear rogue wave formation in the presence of currents, it is very helpful to begin with a ray, or eikonal, approximation for wave evolution in the ocean~\cite{wf,hkd},
\begin{equation}
{d\vec k\over dt} = -{\partial \omega(\vec r,\vec k)\over \partial \vec r}; \ \ \ \ \ \ \ {d\vec r\over dt} = {\partial \omega(\vec r,\vec k)\over \partial \vec k} \,,
\label{eikonal}
\end{equation}
where $\vec r$ is the ray position, $\vec k$ is the wave vector, and $\omega$ is the frequency. For surface gravity waves in deep water, the dispersion relation is
\begin{equation}
\omega(\vec r,\vec k) = \sqrt{g\vert \vec k\vert} + \vec k \cdot \vec U(\vec r) \,,
\label{dispers}
\end{equation}
where $\vec U(\vec r)$ is the current velocity, assumed for simplicity to be time-independent, and $g=9.81$~m/s is the acceleration due to gravity. The validity of the ray approximation depends firstly on the condition $|\vec k|\xi \gg 1$, where $\xi$ is the length scale on which the current field $\vec U(\vec r)$ is varying, physically corresponding to the typical eddy size. This condition is well satisfied in nature, since wave numbers of interest in the deep ocean are normally of order $k\sim 2 \pi/(100 \, {\rm m})$, while the typical eddy size may be $\xi \sim 5\,{\rm km}$ or larger. Secondly, scattering of the waves by currents is assumed to be weak, i.e., the second term in equation~(\ref{dispers}) should be small compared to the free term. This again is well justified since eddy current speeds $|\vec U|$ are normally less than $0.5\, {\rm m/s}$, whereas the wave speed $v=\partial\omega/\partial k\approx \sqrt{g/4k}$ is greater than $5\, {\rm m/s}$. In section~\ref{seclinnumerical} below, we will explicitly compare
 the ray predictions with results obtained by exact integration of the corresponding wave equation.
 
\begin{figure}[ht]
\centerline{\includegraphics[width=2.8in,angle=0]{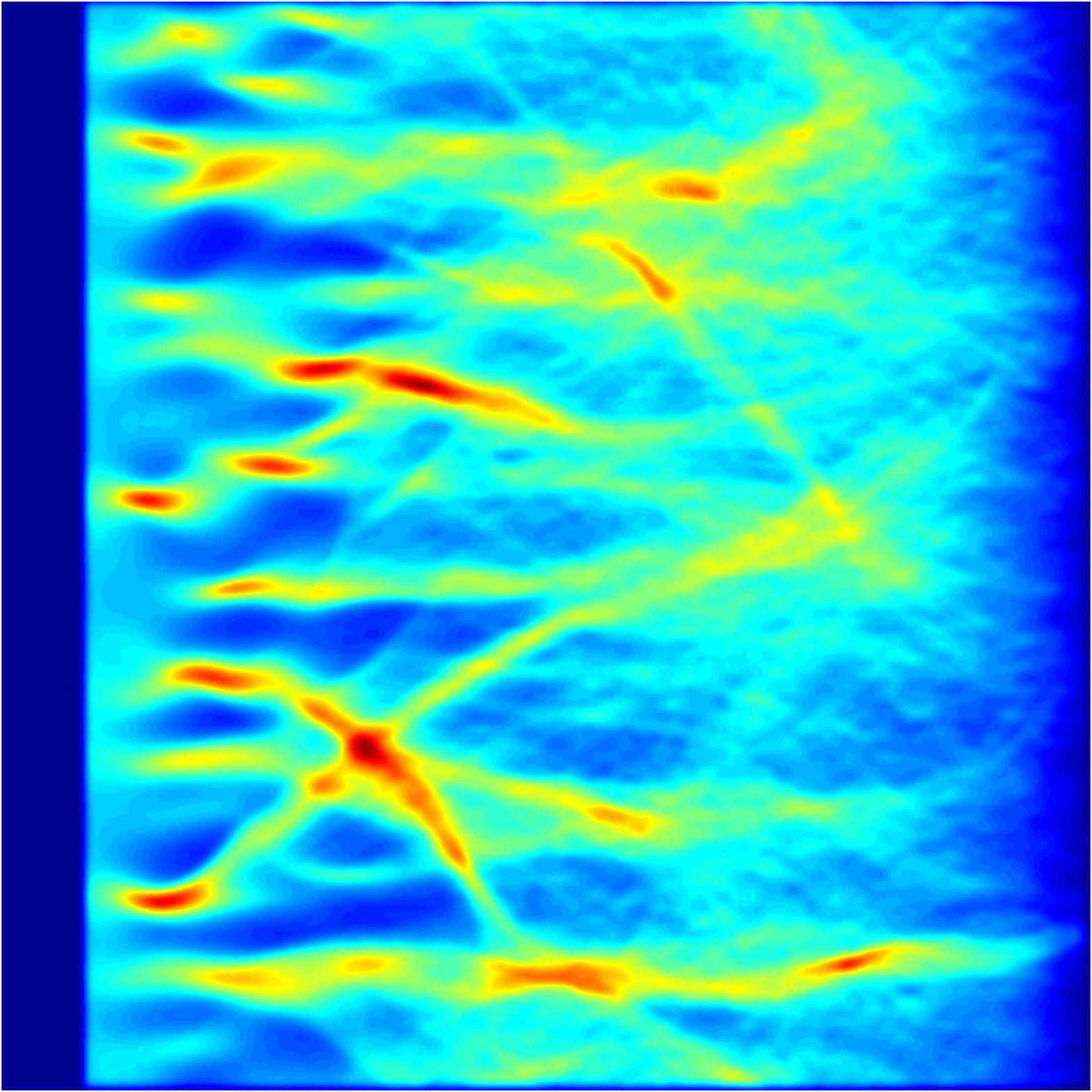}\hskip 0.2in \includegraphics[width=2.8in,angle=0]{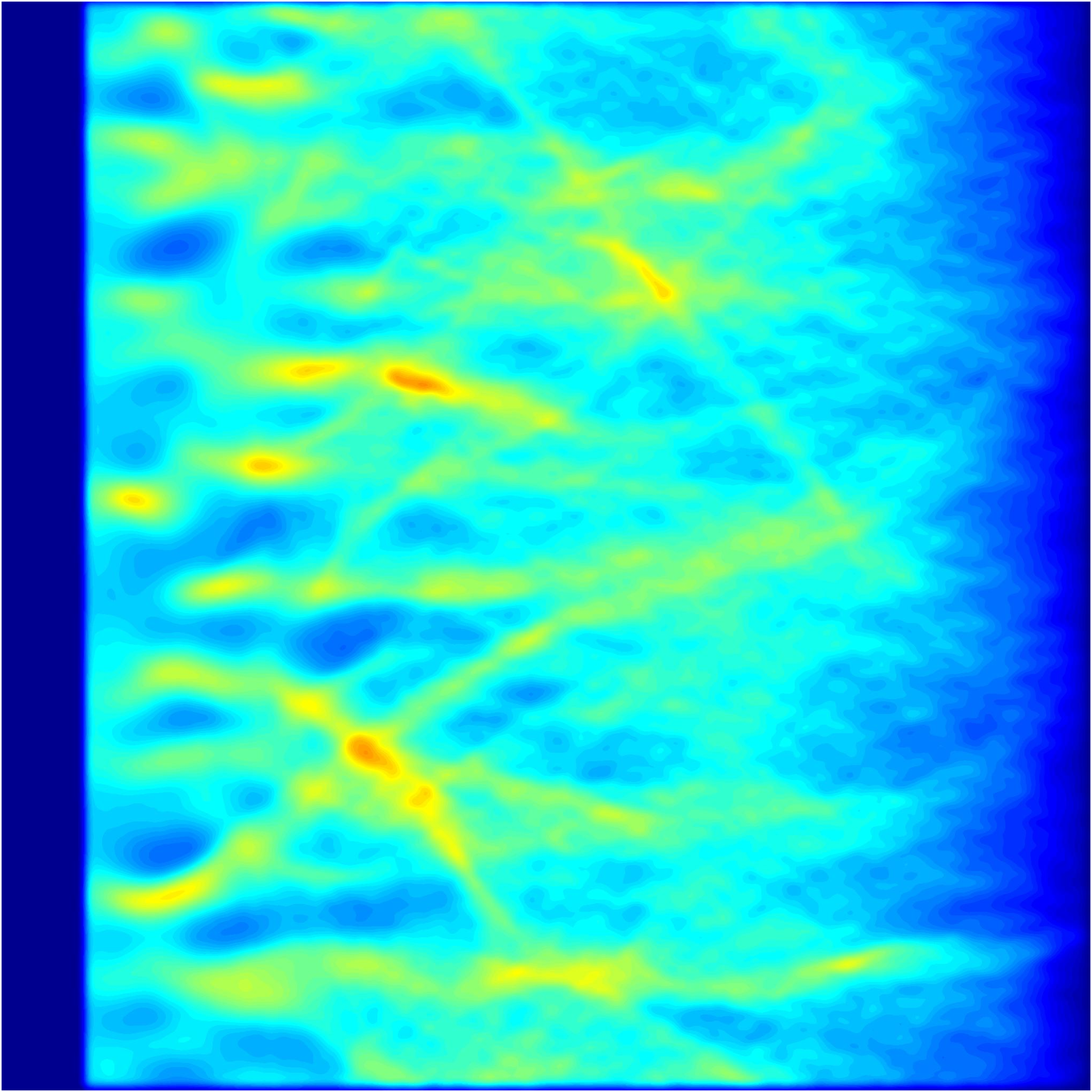}}
%\vskip 0.2in
\caption{A ray density map $I(x,y)$ is calculated for rays moving through a $640$ km by $640$ km random eddy field, with rms eddy current $u_{\rm rms}=0.5$ m/s and eddy correlation length $\xi=20$ km. Here bright regions represent high density. The rays are initially distributed uniformly
along the left edge of each panel, with angular spread $\Delta \theta$ around the $+x$ (rightward) direction, and with frequency $\omega=2 \pi/(10\,{\rm sec})$, corresponding to velocity $v=7.81\,{\rm m/s}$ in the absence of currents. 
The left and right panels illustrate $\Delta \theta=10^\circ$ and $\Delta \theta=20^\circ$, respectively.}
\label{figrayimage}
\end{figure}

In the numerical simulations shown in figure~\ref{figrayimage}, we follow White and Fornberg~\cite{wf} in considering a random incompressible current field in two dimensions,
with zero mean current velocity, generated as
 \begin{equation}
U_x(\vec r) = -{\partial \psi(\vec r)}/{\partial y}\,; \;\;\;\;\;\; U_y(\vec r) = {\partial \psi(\vec r)}/{\partial x}
\end{equation}
from the scalar stream function $\psi(\vec r)$. The stream function itself is Gaussian distributed with Gaussian decay of spatial correlations:
\begin{equation}
\overline{\psi(\vec r)} = 0\,; \;\;\;\;\;\;  \overline{\psi(\vec r)\,\psi(\vec r')} \sim e^{-(\vec r-\vec r')^2/2\xi^2}\,,
\end{equation}
and the overall current strength is described by $u_{\rm rms}^2=\overline{|\vec U(\vec r)|^2}$. The specific choice of a Gaussian distribution for the stream function is made for convenience only. The detailed structure of the individual eddies on the scale $\xi$ has no effect on the final rogue wave statistics as long as the current is weak ($u_{\rm rms} \ll v$), since each ray must travel a distance $\gg \xi$ before being appreciably scattered. Each panel in figure~\ref{figrayimage} represents a $640$ km by $640$ km random eddy field, with rms eddy current $u_{\rm rms}=0.5$ m/s and eddy correlation length $\xi=20$ km.
 
The initial swell, entering from the left in each panel, is characterized by a single frequency $\omega=2\pi/(10\,{\rm sec})$ (and thus a single wave number $k=\omega^2/g=0.04\,{\rm m}^{-1}$ and a single wave speed $v=\sqrt{g/4k}=7.81$ m/s). As discussed in Ref.~\cite{hkd}, within the context of a linear model, a nonzero frequency spread affects rogue wave formation only at second order in the spread $\Delta \omega$, and may be neglected for all practical purposes. In contrast, the {\it angular} spread of the incoming sea is very important in determining rogue wave statistics. In this figure, we consider an initially Gaussian angular distribution $p(\theta) \sim e^{-\theta^2/2(\Delta \theta)^2}$, where $\theta$ is the wave vector direction relative to the mean direction of wave propagation. Here all rays begin at the left edge of each panel, uniformly distributed in the $y$ direction, and the mean direction of wave propagation is rightward. The left and right panels illustrate the behavior for two different values of the initial angular spread $\Delta \theta$.
 
In both panels we observe bright streaks or branches, corresponding to regions of larger than average ray density $I(x,y)$, and thus larger than average wave intensity.
 The branches may be understood by considering briefly the limiting (unphysical) case of a unidirectional initial sea state ($\Delta \theta=0$), corresponding to a single incoming plane wave. In the ray picture, and in the coordinates of figure~\ref{figrayimage}, the initial conditions are in this limit characterized by a one-dimensional phase space manifold $(x, y, k_x, k_y) = (0,y,k,0)$, where $k$ is the fixed wave number, and $y$ varies over all space. As this incoming plane wave travels through the random current field, it undergoes small-angle scattering, with scattering angle $\sim u_{\rm rms}/v$ after traveling one correlation length $\xi$ in the forward direction. Eventually, singularities appear that are characterized in the surface of section map $[y(0),k_y(0)]\to [y(x),k_y(x)]$ by $\delta y(x) / \delta y(0)=0$, i.e., by local focusing of the manifold of initial conditions at a point $(x,y)$.
 
\begin{figure}[ht]
%\centerline{\includegraphics[width=3.8in,angle=0]{figure2.ps}}
\centerline{\includegraphics[width=4in,angle=0]{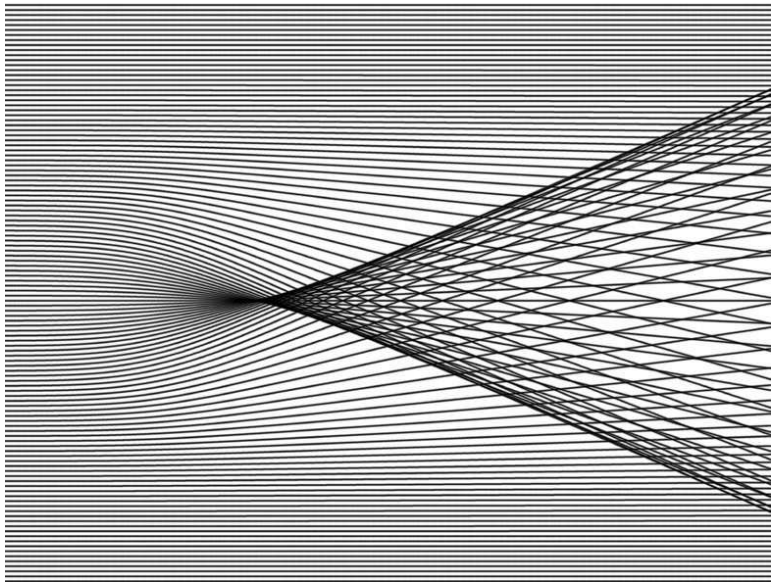}}
%\vskip 0.2in
\caption{A cusp singularity, followed by two branches of fold singularities, is formed as initially
parallel rays pass through a focusing region. The two branches appear because the focal
distance varies with the distance of approach from the center, as in a `bad' lens with strong spherical aberration. After
averaging over incident directions, the singularities will
be softened but not washed away completely~\cite{hkd}}.
\label{figcusp}
\end{figure}

 The currents leading to such a focusing singularity may be thought of as forming a `bad lens.' Whereas a lens without aberration focuses all parallel incoming rays to one point, a bad lens only focuses at each point an infinitesimal neighborhood of nearby rays, so that different neighborhoods get focused at different places as the phase-space manifold evolves forward in $x$, resulting in lines, or branches, of singularities. The typical pattern is an isolated cusp singularity, $\delta^2 x(y) / \delta x(0)^2=0$, followed by two branches of fold singularities, as shown in figure~\ref{figcusp}.

A simple scaling argument~\cite{wf,lkbranch,mfg} shows that the first singularities occur after a median distance $y = L \sim \xi (u_{\rm rms}/v)^{-2/3} \gg \xi$ along the direction of travel.
when the typical ray excursion in the transverse $x$ direction becomes of order $\xi$. Thus, each ray passes through many uncorrelated eddies before a singularity occurs, and a statistical description is well justified. For realistic parameters, $L \sim 100$ km or more is typical. The cusp singularities formed in this way are separated by a typical distance $\xi$ in the transverse direction, and thus the rms deflection angle by the time these singularities appear scales as
\begin{equation}
\delta \theta \sim \xi/L \sim (u_{\rm rms}/v)^{2/3} \,.
\label{delkx}
\end{equation}
We note that the typical deflection angle $\delta \theta$ does not depend on the eddy size but only on the velocity ratio $u_{\rm rms}/v$: faster currents cause larger deflection. For the input parameters used in figure~\ref{figrayimage}, the median distance to the first singularity is $L=7.5\xi=150$~km, and the rms deflection at the point of
singularity is $\delta \theta=18^\circ$.

 \begin{figure}[ht]
\centerline{\includegraphics[width=3.5in,angle=0]{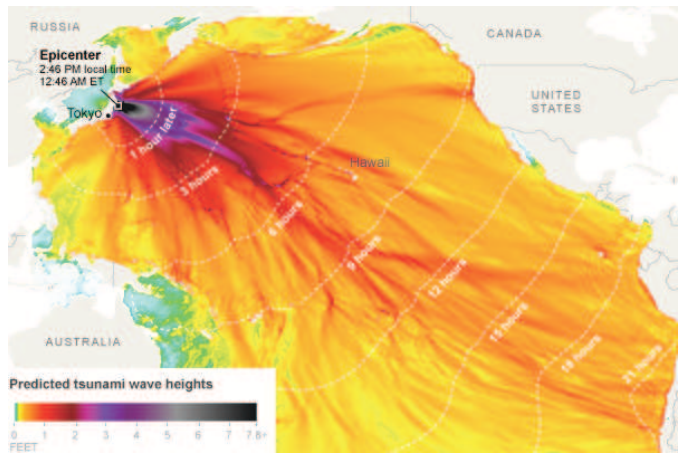}}
%\vskip 0.2in
\caption{Predicted tsunami wave heights from the T\=ohoku earthquake, a 9.0 magnitude undersea earthquake that occurred on March 11, 2011, off the coast of Japan. A branching structure is clearly visible as the waves move outward from the epicenter. (Source: NOAA Center for Tsunami Research.)}
\label{figtsunami}
\end{figure}

Similar phenomenology can give rise to wave focusing and rogue wave formation in shallow water,
where the dispersion relation of equation~(\ref{dispers}) is replaced with $\omega(
{\vec r},{\vec k})=\sqrt{gk \tanh(k h({\vec r}))}$, and varying depth
$h({\vec r})$ takes the place of the varying current $U({\vec r})$ as the
origin of scattering~\cite{tucker}. The same mechanism can lead to amplification of tsunami waves~\cite{berrytsunami,Dobrokhotov} where because of the long wavelength, shallow water equations apply. Fig.~\ref{figtsunami} shows a striking recent example of a predicted tsunami wave height map, in which the branched flow structure is unmistakably present.
More generally,
singularities and branched flow due to focusing in random media have been investigated in contexts as diverse as
electron flow in a two-dimensional electron gas~\cite{2deg}, ocean acoustics~\cite{tomsovic}, twinkling of starlight~\cite{twinkling},
and rain shower activation in turbulent clouds~\cite{rainshower}. Recently, universal
expressions have been obtained describing the branching statistics
for a large class of such systems, and valid at all distances from a source~\cite{mfg}.

 \begin{figure}[ht]
\centerline{\includegraphics[width=3.5in,angle=270]{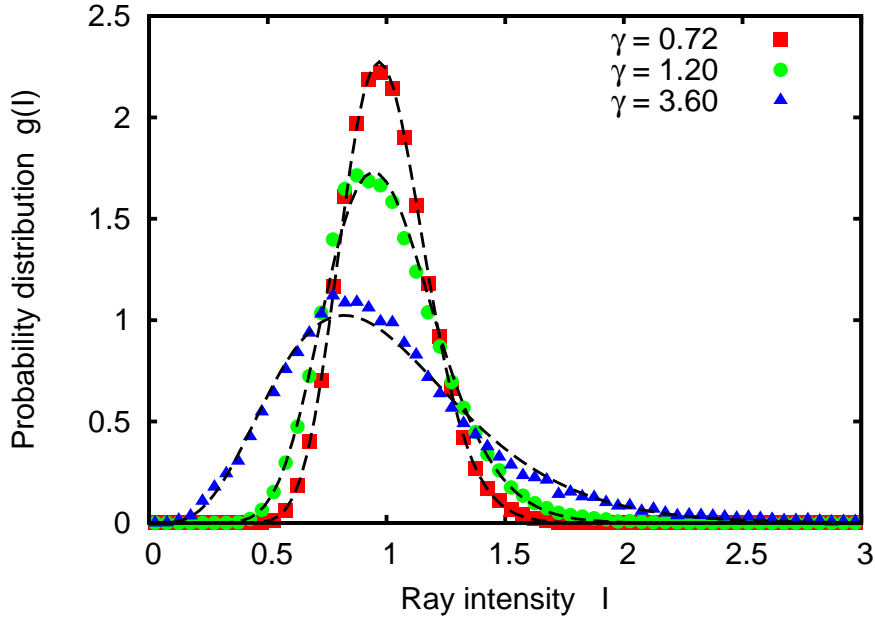}}
%\vskip 0.2in
\caption{The ray density distribution, for an initial sea state of uniform density scattered by a random eddy current field, is shown for several values of the freak index $\gamma$. The input parameters are chosen as in figure~\ref{figrayimage}, with initial angular spread $\Delta \theta=25^\circ$, $15^\circ$, and $5^\circ$ corresponding to freak index $\gamma=0.72$, $1.20$, and $3.60$, respectively. The mean intensity is normalized to unity in each case. The dashed curves are fits to the $\chi^2$ distribution of Eq.~(\ref{chisq}).}
\label{figraychisq}
\end{figure}

For finite initial angular spread $\Delta \theta$, the singularities are softened, and the finite contrast between the peak ray density in the branches and the background intensity is governed for $\Delta \theta \ll 1$ and $\delta \theta \ll 1$ by the ratio
\begin{equation}
\gamma =  {\delta \theta \over \Delta \theta}  \sim {(u_{\rm rms}/v)^{2/3} \over \Delta \theta}   \,,
\label{gammadef}
\end{equation}
which we refer to as the freak index~\cite{hkd}. Of particular interest is the regime of small $\gamma$, where the scattering characterized by $\delta \theta$ is weak compared to the initial angular spread $\Delta \theta$ of the incoming sea. In this limit, the scattering produces only small perturbations of order $\gamma^{-1}$ in the ray density $I(x,y)$, in units where the initial (uniform) density is $I_0=1$~\cite{hkd}. Specifically, 
as seen in figure~\ref{figraychisq},
the distribution of ray intensities in this regime may be well described
by a $\chi^2$ distribution~\cite{microwave},
\begin{equation}
g(I)=\chi_N^2(I)=\left(\frac{N}{2}\right)^{\frac{N}{2}}\frac{I^{\frac{N}{2}-1}}{\Gamma\left(\frac{N}{2}\right)}
  e^{-NI/2} \,,
\label{chisq}
\end{equation}
where the number of degrees of freedom $N$ scales with the freak index as $\gamma^{-2}$. The proportionality constant may be obtained numerically by a fit to the data:
\begin{equation}
N={\alpha \over \gamma^2}={45 \over \gamma^2} \,.
\label{nval}
\end{equation}
In the limit $\gamma \to 0$ associated with zero current, we have $N \to \infty$, and we recover as expected the uniform ray density distribution
$g(I)=\delta(I-1)$.

\subsection{Implications for Wave Statistics}
\label{linwavestat}

In the Longuet-Higgins random seas model~\cite{lh}, the sea surface elevation above the average elevation is given by ${\rm Re} \, \zeta(x,y,t)$, where $\zeta$ is a random superposition of many plane waves with differing directions and frequencies. By the central limit theorem, $\zeta$ is distributed as a complex Gaussian random variable with standard deviation $\sigma$. Furthermore, for a narrow-banded spectrum ($\delta \omega \ll \omega$) the wave crest height $H$ is equal to the wave function amplitude $|\zeta|$, and the probability of encountering a wave crest of height $H$ or larger is
\begin{equation}
P_{\rm Rayleigh}(H) = e^{-H^2/2 \sigma^2} \,.
\label{prayleigh}
\end{equation}
Due to an exact symmetry between crests and troughs in a linear wave model, a crest height of $H$ corresponds to a wave height (crest to trough) of $2H$. Conventionally, a rogue wave is defined as $2H \ge 2.2\,{\rm SWH}$, where the significant wave height ${\rm SWH}$ is the average of the largest one third of wave heights in a time series, or approximately ${\rm SWH} \approx 4.0 \sigma$. Thus the condition for a rogue wave is $H \ge 4.4 \sigma$, and the random seas model predicts such waves to occur with probability $P_{\rm Rayleigh}(4.4\sigma) =6.3 \cdot 10^{-5}$. Similarly, extreme rogue waves may be defined by the condition $2H \ge 3\,{\rm SWH}$ or $H \ge 6.0\sigma$, and these are predicted to occur with probability $P_{\rm Rayleigh}(6.0\sigma) =1.5 \cdot 10^{-8}$ within the random seas model. As discussed in section~\ref{secintro}, the random seas model greatly underestimates the occurrence probability of extreme waves, when compared with observational data~\cite{dankert}.

What are the implications of scattering by currents, as discussed in section~\ref{secrayinten}, on the wave height statistics? Within the regime of validity of the ray approximation, we have at any spatial point $(x,y)$ correspondence between the ray density $I(x,y)$ and the wave intensity $H^2=|\zeta(x,y,t)|^2$, averaged over time. Thus, in contrast with the original Longuet-Higgins model, the time-averaged wave intensity is not uniform over all space but instead exhibits ``hot spots'' and ``cold spots'' associated with focusing and defocusing in the corresponding ray equations. At each point in space (assuming of course that the currents are stationary), the central limit theorem and thus the Rayleigh distribution still apply, and we have
\begin{equation}
P_{(x,y)}(H) = e^{-H^2/2\sigma^2 I(x,y)} \,,
\label{plocal}
\end{equation}
where $I(x,y)$ is the local ray density, normalized so that the spatial average is unity, and
$\sigma^2$ is the variance of the surface elevation in the incoming sea state, before scattering by currents. This is the situation a ship experiences at a given position.

Now averaging over space, or over an ensemble of random eddy fields with a given rms current speed, we obtain a total cumulative height distribution 
\begin{equation}
P_{\rm total}(H) = \int_0^\infty dI\,  g(I) \, e^{-H^2/2\sigma^2 I} \,.
\label{ptotal}
\end{equation}
In equation~(\ref{ptotal}), the full cumulative distribution of wave heights for a given sea state has been expressed as a convolution of two factors: (i) the local density distribution $g(I)$, which can be extracted from the ray dynamics, and (ii) the universal Longuet-Higgins distribution of wave heights for a given local density. Similar decompositions of chaotic wave function statistics into non-universal and universal components have found broad applicability in quantum chaos, including for example in the theory of scars~\cite{scar,baecker}. In the context of rogue waves, a similar approach was adopted by Regev {\it et al.} to study wave statistics in a one-dimensional inhomogeneous sea,
where the inhomogeneity arises from the interaction of an initially homogeneous sea with a 
(deterministic) long swell~\cite{regev}.

Using the previously obtained ray density distribution in the presence of currents, equation~(\ref{chisq}), we obtain the K-distribution~\cite{kdistr}
\begin{equation}
P_{\rm total}(H)=2
\frac{\;\;\left({\sqrt{N} H/2\sigma}\right)^{\frac{N}{2}}}{\Gamma(N/2)}
 K_{N/2}\left(\sqrt{N} H \sigma\right)\,,
 \label{kbess}
\end{equation}
where $K_n(y)$ is a modified Bessel function.

Defining the dimensionless variable $x=2H/{\rm SWH} \approx 2H/(4 \sigma)$, so that a rogue wave is given by $x=2.2$ and an extreme rogue wave by $x=3.0$, we find
the probability of a wave height exceeding $x$ significant wave heights:
\begin{equation}
P_{\rm total}(x)=2
\frac{\;\;\left(\sqrt{N}x\right)^{\frac{N}{2}}}{\Gamma(N/2)}
 K_{N/2}\left(2\sqrt{N}x\right)\,,
 \label{kbess2}
\end{equation}
to be compared with the random seas prediction
\begin{equation}
P_{\rm Rayleigh}(x)=e^{-2 x^2}
\label{prayleighx}
\end{equation} in the same dimensionless units.
We recall that $N$ in equation (\ref{kbess}) or (\ref{kbess2})
is a function of the freak index $\gamma$, as given by equation (\ref{nval}).

To examine the predicted enhancement in the probability of rogue wave formation, as compared with random seas model (\ref{prayleigh}), we may consider two limiting cases. Keeping the wave height of interest fixed, and taking the limit $\gamma \to 0$, i.e. $N \to \infty$, we obtain the perturbative result
\begin{eqnarray}
P_{\rm perturb}(x)&=&\left[ 1+\frac{4}{N}(x^4-x^2)\right]P_{\rm Rayleigh}(x) \\
&=& \left[ 1+\frac{4 \gamma^2}{b}(x^4-x^2)\right]P_{\rm Rayleigh}(x) \,,
\label{pperturb}
\end{eqnarray}
valid for $x^4 \ll N$, or equivalently $x^2 \gamma \ll 1$. Thus, in the limit of small freak index, the distribution reduces, as expected, to the prediction of the random seas model.
Analogous perturbative corrections appear for quantum wave function intensity distributions in the presence of weak disorder or weak scarring by periodic orbits~\cite{mirlin,damborsky}.

Much more dramatic enhancement is observed if we consider the tail of the intensity distribution ($x \to \infty$) for a given set of sea conditions (fixed $\gamma$ or $N$). Then for $x \gg N^{3/2}$, or equivalently $x \gamma^3 \gg 1$, we  obtain the asymptotic form
\begin{eqnarray}
P_{\rm asymptotic}(x)&=&
 \sqrt{\pi}
\frac{\;\;\left(\sqrt{N}x\right)^{\frac{N-1}{2}}}{\Gamma(N/2)}
 e^{-2x \sqrt{N}}
 \nonumber \\
 &=&
 \sqrt{\pi}
 \frac{\;\;\left(\sqrt{N}x\right)^{\frac{N-1}{2}}}{\Gamma(N/2)}
 e^{2x(x- \sqrt{N})} P_{\rm Rayleigh}(x)
 \,,
 \label{asympt}
\end{eqnarray}
i.e., the probability enhancement over the random seas model is manifestly super-exponential in the wave height $x$.

Predicted enhancements in the probability of rogue wave and extreme rogue wave formation, based on equations (\ref{kbess2}) and (\ref{prayleighx}), are shown in table~\ref{enhancement}. We notice in particular that enhancements of an order of magnitude or more are predicted in the extreme tail, even for moderate values of the input parameters, corresponding to $\gamma \sim 0.72\,-\,1.2$ or $N \sim 30 \, - \, 85$.

\begin{table}[ht]
\begin{center}
%\begin{tabular*}{0.80\textwidth}{@{\extracolsep{\fill}} | c | c | c | c |c |}
\begin{tabular}{| c | c | c | c | c |}
  \hline
  \;\;\;\;\;$\Delta \theta$\;\;\;\;\; & \;\;\;\;\;$\gamma$\;\;\;\;\; & \;\;\;\;\;$N$\;\;\;\;\; & \;\;\;\;\;$E(2.2)$\;\;\;\;\; & \;\;\;\;\;$E(3.0)$\;\;\;\;\; \\
  \hline
   5 & 3.6 & 3.46 & 57 & 16800 \\
  10 & 1.8 & 13.9 & 10.4 & 570 \\
  15 & 1.2 & 31.2 & 4.3 & 76 \\
  20 & 0.90 & 55.4 & 2.7 & 22 \\ 
  25 & 0.72 & 86.6 & 2.0 & 9.8 \\
  30 & 0.60 & 125 & 1.7 & 5.7 \\
  \hline
\end{tabular}
\end{center}
\caption{The $N$ parameter of equation~(\ref{kbess2}), and the associated enhancement in the probability of rogue wave formation (wave height $2H=2.2\, {\rm SWH}$) as well as the enhancement of the probability of extreme rogue wave formation
(wave height $2H=3.0\, {\rm SWH}$) are calculated for several values of the incoming angular spread $\Delta \theta$
using equations~(\ref{gammadef}), (\ref{nval}), and (\ref{kbess2}). Here $E(x)=P_{\rm total}(x)/P_{\rm Rayleigh}(x)$. In all cases we fix the rms current speed $u_{\rm rms}=0.5$~m/s and mean wave speed $v=7.8$~m/s, so $\delta \theta=18^\circ$.
}
\label{enhancement}
\end{table}

\subsection{Numerical Results for Linear Wave Equation}
\label{seclinnumerical}

The theoretical predictions of equation~(\ref{kbess2}) are based on several approximations, including the assumption of local Rayleigh statistics. To see whether the approximations we have made are valid, we compare the theoretical predictions with direct numerical integration of the current-modified linear Schr\"odinger equation, which is obtained from the third-order current-modified nonlinear Schr\"odinger (CNLS) equation~\cite{stocker} by setting the nonlinear term to zero. CNLS governs the modulations of weakly nonlinear water waves around a mean frequency and mean wave vector, incorporating the effect of currents, and is presented in full in Sec.~\ref{secnlse} below. In dimensionless variables, the linear equation for the wave envelope describing the wave modulations is expressed as~\cite{stocker}
\begin{equation}
iA_T -\frac{1}{8}A_{XX}+ \frac{1}{4}A_{YY}-k_0 U_x A=0 \,.
\label{cnls}
\end{equation}
Here $A(X,Y,T)$ is the wave envelope, defined by separating out the carrier wave propagating
with mean wave vector ${\vec k}= k_0 {\hat x}$,
\begin{eqnarray}
\zeta(X,Y,T)&=&k_0 A(X,Y,T)e^{i k_0 x-i\sqrt{gk_0}t} \nonumber \\
&=& k_0 A(X,Y,T)e^{i X-iT/2}\,,
\end{eqnarray}
and 
\begin{equation}
(X,Y,T)=(k_0 x-\frac{1}2{}\sqrt{gk}t,k_0 y,\sqrt{g k_0}t)
\label{dimspacetime}
\end{equation}
are dimensionless space and time coordinates. We also note that Eq.~(\ref{cnls}) may be obtained directly from the dispersion relation (\ref{dispers}), by expanding $\omega$ and $\vec k$ around $\omega_0$ and $k_0 {\hat x}$, respectively.

\begin{figure}[ht]\centerline{\includegraphics[width=3.0in,angle=270]{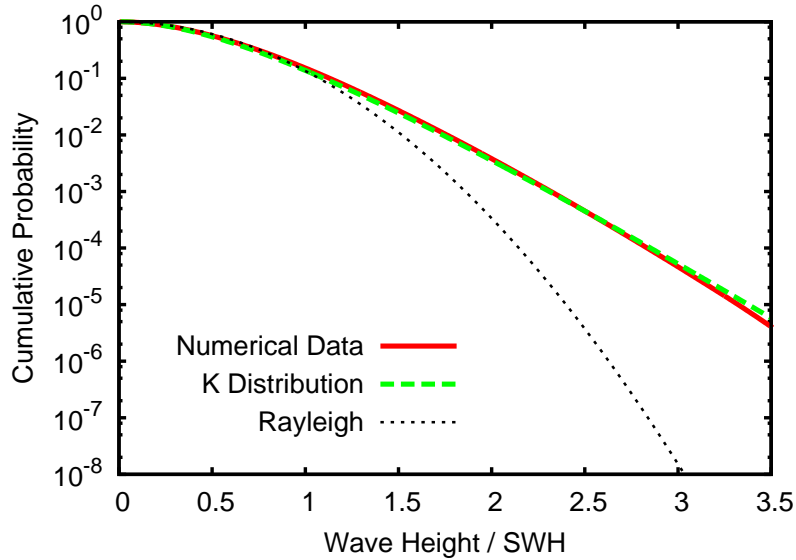}}
%\vskip 0.2in
\caption{The probability of exceeding wave height $2H$, in units of the significant wave height SWH, is shown for an incoming wave speed $v=7.8$~m/s, incoming angular spread $\Delta \theta=5.7^\circ$, and rms current speed $u_{\rm rms}=0.5$~m/s. The solid curve shows the results of a numerical simulation performed on a $20$~km by $40$~km field, with typical eddy size $\xi=800$~m, while the dashed curve represents equation~(\ref{kbess2}) with $N=6.8$. The Rayleigh (random seas) prediction of equation (\ref{prayleighx}) is shown for comparison.
}
\label{figkdistr}
\end{figure}

The calculations are performed on a rectangular field measuring 40 km along the mean direction of propagation and 20 km in the transverse direction, with typical eddy size $\xi=800$~m. (We note that a very small value for the eddy size is chosen to maximize the statistics collected; this is also a ``worst case'' scenario for the theory, as the ray approximation is expected to work ever better as the ratio of eddy size to the wavelength increases.) Equation~(\ref{cnls}) is integrated numerically using a split-operator Fourier transform method~\cite{weidman}, on a 1024 by 512 grid. The incoming wave is a random superposition of a large number of monochromatic waves with 
directions uniformly distributed around the mean direction $\theta=0$ with standard deviation $\Delta \theta$. Without loss of generality, the incoming wave number is fixed at $k_0=2 \pi/(156\,{\rm m})$, corresponding to a frequency $\omega_0=\sqrt{gk_0}=2 \pi/ (10\,{\rm sec})$ and a group velocity $v=7.81$~m/s.
%The steepness $k_0 \overline{H}$ is adjusted by varying the mean height $H_0$ of the incoming sea.
Each run simulates wave evolution for $5 \cdot 10^5$~sec or $5\cdot 10^4$ wave periods, sufficient for the wave height statistics to converge.

The results for $\Delta \theta=5.7^\circ$, corresponding to a very large freak index $\gamma=3.15$, are shown in figure~\ref{figkdistr}. The results are compared both with the theoretical prediction of equation~(\ref{kbess2}) (here $N=6.8$) and with the baseline Rayleigh distribution of equation~(\ref{prayleighx}). This is an extreme scenario, in which the occurrence probability of extreme rogue waves ($3$ times the significant wave height) is enhanced by more than three orders of magnitude. Even better agreement with the theoretical model of equation~(\ref{kbess2}) obtains for more moderate values of $\gamma$, corresponding to larger $N$.

\begin{figure}[ht]\centerline{\includegraphics[width=3.0in,angle=270]{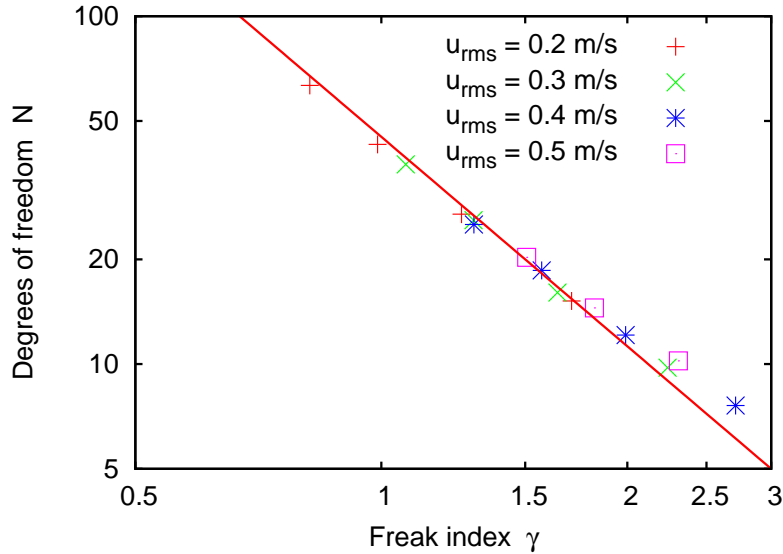}}
%\vskip 0.2in
\caption{The wave height distribution for a random incoming sea scattered by random currents is obtained numerically for four values of the rms current speed $u_{\rm rms}$ and four values of the incoming angular spread $\Delta \theta$. In each case, a fit to the K-distribution (equation~(\ref{kbess})) yields the number-of-degrees-of-freedom parameter $N$, (describing deviations from Rayleigh statistics), which is plotted as a function of the freak index $\gamma$ (defined in equation~(\ref{gammadef})). As in the previous figures, the wave speed is fixed at $v=7.81$~m/s. The solid line is the theoretical prediction of equation~(\ref{nval}).
}
\label{figlinscaling}
\end{figure}

In figure~\ref{figlinscaling} we repeat the numerical simulation for four different values of the incoming angular spread $\Delta \theta$ and four different values of the rms current speed $u_{\rm rms}$. In each case, the numerically obtained wave height distribution is fit to a K-distribution (equation~(\ref{kbess})), and the resulting value of $N$ (which fully describes the strength of deviations from Longuet-Higgins statistics) is plotted as a function of the freak index $\gamma$. Excellent agreement is observed with the power-law prediction of equation~(\ref{nval}) all the way up to $\gamma \approx 2$ (corresponding to $N \approx 10$), even though the analytic prediction was obtained in a small-$\gamma$ approximation. The regime in which the analytic formula (\ref{nval}) works well includes most conditions likely to be found in nature (e.g., all but the first row of table~\ref{enhancement}). Referring again to table~\ref{enhancement}, we observe that the theory accurately describes enhancements of up to three orders of magnitude in the formation probability of extreme rogue waves. Modest deviations from the analytic formula are observed numerically at very large values of $\gamma$ (corresponding to even larger enhancements).

\subsection{Experimental Demonstration of Linear Rogue Wave Formation}

\begin{figure}[ht]\centerline{\includegraphics[width=3.0in,angle=0]{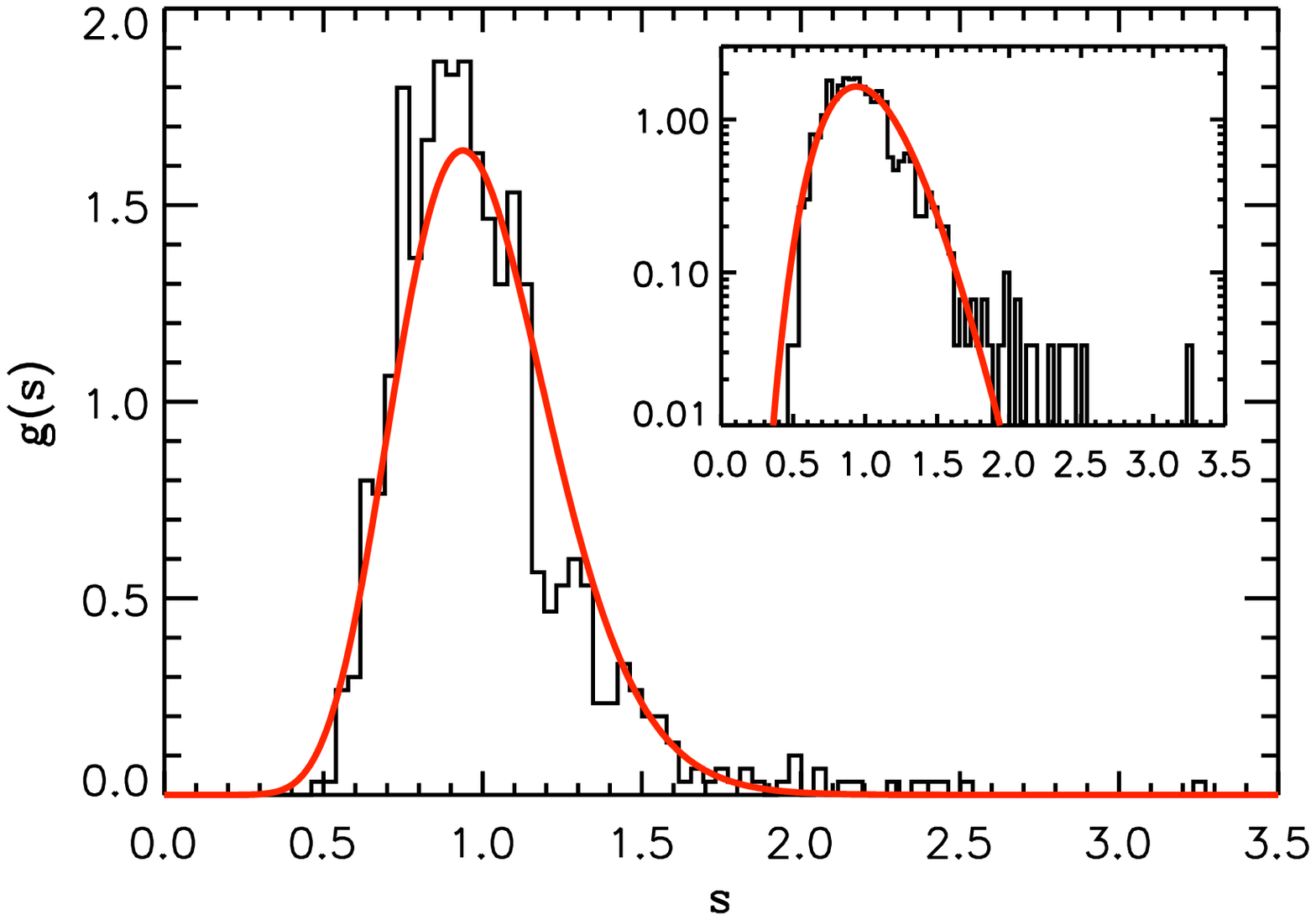} \includegraphics[width=3.0in,angle=0]{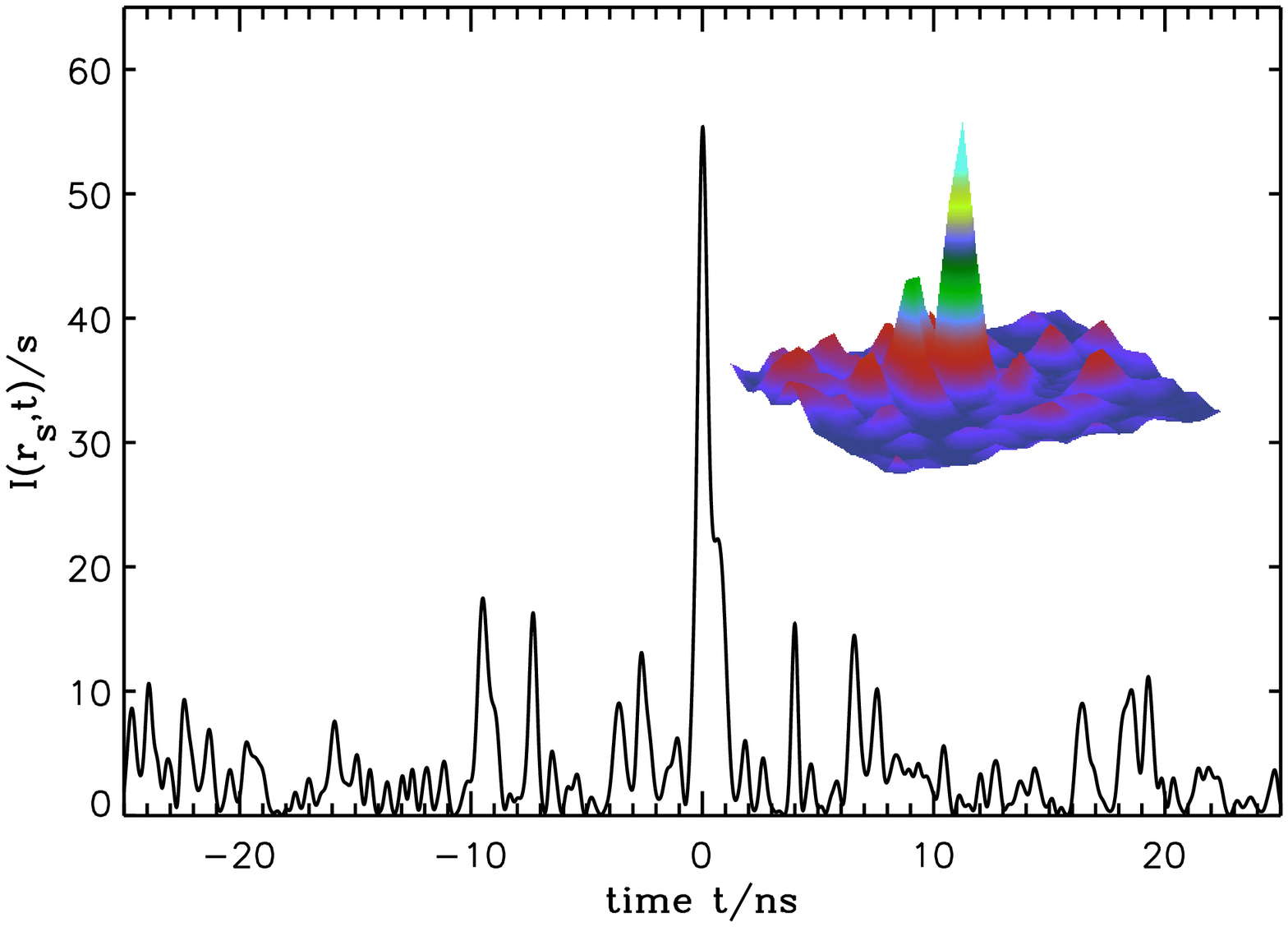}}
%\vskip 0.2in
\caption{The left panel shows the distribution of the experimentally measured time-averaged microwave intensity $s$, found for $780$ positions in the random wave field, is compared with the $\chi^2$ distribution of 
equation~(\ref{chisq}), with $N=32$. The inset shows the same data on a logarithmic scale. The right panel shows a time series of the wave intensity at a single point shows a ``rogue wave'' event. The inset shows a snapshot of the wave intensity field near this point at the moment of the extreme event.
}
\label{figmicrowave}
\end{figure}

Direct experimental verification in the ocean of the statistical predictions made analytically in section~\ref{linwavestat} and confirmed numerically in section~\ref{seclinnumerical} is obviously highly desirable. Unfortunately no observational data set exists at this time that would allow the tail of the wave height distribution to be studied as a function of the freak index $\gamma$, i.e., as a function of the rms current speed $u_{\rm rms}$ and of the angular spread $\Delta \theta$. Recently, however, experiments in open quasi-two-dimensional microwave cavities with disorder~\cite{microwave} have found a strong enhancement in the occurrence probability of high-amplitude waves, which may be interpreted as ``rogue waves'' in this analog system. In the microwave system, randomly placed brass cones play the role of random ocean currents, and a movable source antenna enables incoming waves to arrive from different directions. A movable drain antenna acts a weak probe, and allows for a spatial mapping of the wave fields within the scattering arrangement. A great advantage of the microwave system is that the electromagnetic wave equation is linear, so that the observed enhancement in the tail of the wave height distribution may serve in principle as a direct experimental test of the theory developed in the previous sections.

In the left panel of figure~\ref{figmicrowave}, the time-averaged wave intensity $s$ is found for different positions of the probe, and the probability distribution $g(s)$ is shown. Most of the distribution is well described by the $\chi^2$ distribution of equation~(\ref{chisq}). We note that in the absence of disorder, the time-averaged intensity would be position-independent and $g(s)$ would reduce to $\delta(s-1)$ ($N \to \infty$ in equation~(\ref{chisq})). The inset in the left panel shows additional rare events in the far tail,that are not described by the $\chi^2$ distribution~\cite{microwave}. The right panel in figure~\ref{figmicrowave} shows time series data of the wave intensity at a single point, including an extreme event observed in the experiment, and the inset shows a snapshot of the wave intensity in the region at the moment corresponding to this extreme event. The event presented here has wave height $2H=5.3$~SWH, and events of this magnitude of greater are observed with probability $1.3 \times 10^{-9}$ in the experiment, which is an enhancement of 15 orders of magnitude compared to the Rayleigh distribution.

These results confirm that linear scattering is a sufficient mechanism for a large enhancement in the tail of the wave height distribution, even when nonlinearity is entirely absent from the physical system being studied.

\section{Nonlinear Wave Model}

We have already seen (e.g., in table~\ref{enhancement}) that under physically realistic sea conditions, linear wave dynamics, with nonlinearity only in the corresponding ray equations, are sufficient to 
enhance the incidence of extreme rogue waves by several orders of magnitude. At the same time, the true equations for ocean wave evolution are certainly nonlinear, and furthermore the nonlinear terms, which scale as powers of the wave height, manifestly become ever more important in the tail of the wave height distribution. Thus, a fully quantitative theory of rogue wave statistics must necessarily include nonlinear effects, which we address in the following.

\subsection{Nonlinear Schr\"odinger Equation}

\label{secnlse}

The original Nonlinear Schr\"odinger Equation (NLSE) for surface gravity waves in deep water was derived by Zakharov using a spectral method~\cite{zakharov}, and is valid
to third order in the steepness $\varepsilon =k_0 \overline{H}$, where $\overline{H}$ is the
mean wave height. Subsequently, the NLSE was extended to fourth order in $\varepsilon$
by Dysthe~\cite{dysthe} and then to higher order in the bandwidth $\Delta \omega/\omega$ by Trulsen and Dysthe~\cite{trulsen}. The Trulsen-Dysthe equations include frequency downshifting~\cite{trulsendownshift}, the experimentally observed reduction in average frequency over time~\cite{lake}; however the physics of frequency downshifting may not yet be fully understood~\cite{segur}.

In our simulations we implement the current-modified
$O(\varepsilon^4)$ NLSE, as derived by Stocker and Peregrine in dimensionless form~\cite{stocker}:
\begin{eqnarray}\label{cnls4}
& &i{B}_T -\frac{1}{8}({B}_{{X}{X}}-2{B}_{{Y}{Y}})-\frac{1}{2}{B}|{B}| ^2-{B}\Phi_{c{X}}\nonumber=  \frac{i}{16} ({B}_{{X}{X}{X}}-6{B}_{{Y}{Y}{X}}) \nonumber \\ & &+\bar{\Phi}_{{X}}{B}+\frac{i}{4}{B}({B}{B^*}_{{X}}-6{B^*}{B}_{{X}})
+i(\frac{1}{2}\Phi_{c{X}T}-\Phi_{cZ}){B}-i\bar{\nabla}_h\Phi_c\cdot\bar{\nabla}_h{B} \,,
\end{eqnarray}
where the the linear and third-order terms are collected on the left hand side of equation~(\ref{cnls4}). 
Here $\bar{\Phi}$, $\Phi_c$, and $B$ represent the mean flow, surface current, and oscillatory parts, respectively of
the velocity potential $\phi$:
\begin{equation}\label{vpexpand}
\phi=\sqrt{\frac{g}{k_0^3}}\left[\bar{\Phi}+\Phi_c+ \frac{1}{2}\left( Be^{k_0z+i\theta}+B_2e^{2(k_0z+i\theta)}+{\rm c.c.} \right)\right]\,,
\end{equation}
where the second-harmonic term $B_2$ is function of $B$ and its derivatives,
$({X},{Y},T)$ are dimensionless space and time coordinates defined previously in equation~(\ref{dimspacetime}),
and $\theta=k_0x-\sqrt{gk_0}t={X}-T/2$ is the phase. The surface elevation, which is the quantity of interest
for our purposes, is similarly expanded as
 \begin{equation}
\zeta={k_0}^{-1}\left[\bar{\zeta}+\zeta_c+ \frac{1}{2}\left( A^{i\theta}+A_2e^{2i\theta}+A_3e^{3i\theta}+{\rm c.c.} \right)\right]\,,
\label{Aexpand}
\end{equation}
where the expansion coefficients may be obtained from the velocity potential as
\begin{eqnarray}
A&=&i B+\frac{1}{2k_0} B_x+\frac{i}{8 k_0^2}(B_{xx}-2B_{yy})+\frac{i}{8} B|B|^2 \nonumber \\
A_2&=&-\frac{1}{2}B^2 +\frac{i}{k_0}BB_x \\
A_3&=&-\frac{3i}{8} B^3 \nonumber \,.
\end{eqnarray}
***Here both B and A are of order $\varepsilon$, and proportion to $\varepsilon$. By changing the magnitude of B or A in the incoming wave, we can set steepness to different value. 

In the simulatin, for the simplicity, we works in the frame of reference moving with the velocity $v_0=(c_o+U_o,V_o)$, so $\bar{\Phi}$ and $\Phi_c$ in equation~(\ref{vpexpand}) is zero. The incoming wave is a random superposition of a large number of monochromatic waves with different frequencies and propagating directions. Thus the initial wave could be prepared analytically, as a linear summation of a large number of plain wave,
\begin{equation}\label{eq:iniwave}
\psi(\vec{r},t)=\sum_1^N\phi_i=\sum_1^NA_ie^{i\vec{k_i}\cdot\vec{r}} 
\end{equation}
where $\vec{k_i}$ is the random wave vector for each monochromatic wave. For our setup, the wave vector can be expressed as
\begin{equation}\label{eq:wavenumber}
\vec k=(k_0+k')\cdot (cos\theta'\,\vec x+sin\theta'\,\vec y)
\end{equation}
where $k'$ is a random variation in wave number follows a Gaussian distribution whose half height width is $\Delta k$, and $\theta'$ is the angular spread which is a normal distribution with stand deviation $\Delta \theta$.

In the following examples, equation~(\ref{cnls4}) is integrated numerically with the current set to zero, in order to investigate systematically and quantitatively the effect of nonlinear focusing. In nature, the interplay between linear and nonlinear mechanisms is also of great interest, and may give rise to even stronger enhancement in the probability of rogue wave occurrence than either effect individually, as demonstrated below in section~\ref{seccombined}~(see also~\cite{janssenherbers,yingkaplan}).

\subsection{Height Distribution}
\label{secnlheight}

As in the linear case, the split-operator Fourier transform method is used to integrate equation~(\ref{cnls4}) numerically. The rectangular field measuring 20 km along the mean direction of propagation and 10 km in the transverse direction is discretized using a 1024 by 512 grid. The incoming state is a random superposition of plane waves with wave numbers normally distributed around $k_0$ with standard deviation $\Delta k$, and
directions uniformly distributed around the mean direction $\theta=0$ with standard deviation $\Delta \theta$. Without loss of generality we fix the mean incoming wave number at $k_0=2 \pi/(156\,{\rm m})$, as in section~\ref{seclinear}. The steepness $k_0 \overline{H}$ is adjusted by varying the mean height $\overline{H}$ of the incoming sea. Each run simulates wave evolution for $4 \cdot 10^6$~sec or $4\cdot 10^5$ wave periods.

Typical results are represented by solid curves in figure~\ref{fignldistr}, where we fix $\Delta k/k_0=0.1$ and $\Delta \theta=2.6^\circ$ (as
we will see below, the values of $\Delta \theta$ required to see very strong effects from nonlinear focusing are typically smaller than those needed to observe significant deviations from Rayleigh by linear scattering). The cumulative probability distribution of the wave height $2H$, in units of the significant wave height SWH, is shown for four nonzero values of the wave steepness $\varepsilon$. As expected, the Rayleigh probability distribution of equation~(\ref{prayleighx}) is recovered
in the limit $\varepsilon \to 0$, and ever stronger enhancement in the tail is observed as the steepness of the incoming sea increases. The occurrence probability of extreme rogue waves, $2H/{\rm SWH}=3.0$, is enhanced by one to three orders of magnitude for the parameters shown.

To understand the functional form of the distributions in figure~\ref{fignldistr}, we again make use of the local Rayleigh approximation discussed above in section~\ref{linwavestat}. Here the wave height distribution is given locally in space and time by a Rayleigh distribution around the local mean height (corresponding to a locally random superposition of plane waves), while the local mean height itself varies slowly on the scale of the mean wavelength and mean period. This approximation is well justified, since the envelope $A(X,Y,T)$ in equation~(\ref{Aexpand}) is slowly varying for $\Delta k/k_0 \ll 1$ and $\Delta \theta \ll 1$, while the higher harmonics $A_2(X,Y,T)$ and $A_3(X,Y,T)$ are suppressed by factors of $\varepsilon$ and $\varepsilon^2$, respectively. Taking the local mean intensity to be $\chi^2$ distributed, and convolving the $\chi^2$ distribution of the mean intensity with the Rayleigh distribution around the mean intensity, we obtain as in the linear case a K-distribution (\ref{kbess2}) for the total distribution of wave heights.

\begin{figure}[ht]
\centerline{\includegraphics[width=4in,angle=0]{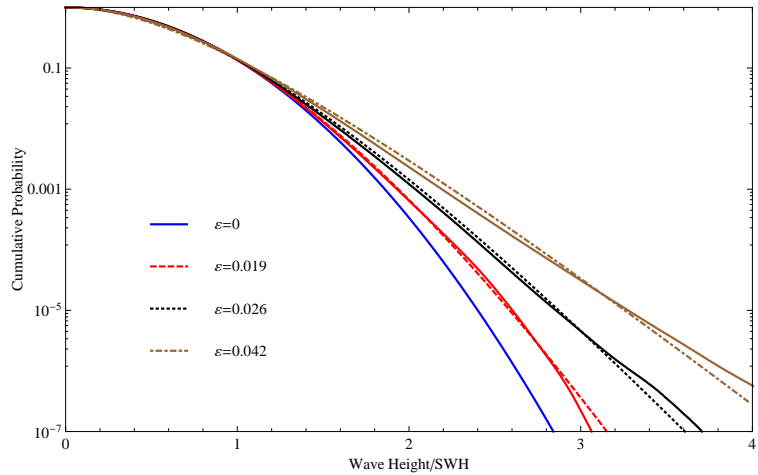}}
%\vskip 0.2in
\caption{The distribution of wave heights, in units of the significant wave height, is calculated for three nonzero values of the steepness $\varepsilon$ (upper three solid curves), and compared with the random seas model of equation~(\ref{prayleighx}) (lowest solid curve). In each case, the dashed or dotted curve is a best fit to the K-distribution of equation~(\ref{kbess2}). Here the we fix the angular spread $\Delta \theta=2.6^\circ$ and wave number spread $\Delta k/k_0=0.1$ of the incoming sea. 
}
\label{fignldistr}
\end{figure}

In figure~\ref{fignldistr}, each data set is fit to the K-distribution of equation~(\ref{kbess2}), arising from the local Rayleigh approximation. We see that the fits, indicated by dashed and dotted lines, perform adequately for probabilities down to $10^{-6}$, where statistical noise begins to dominate. In particular, we clearly observe the crossover between the Gaussian behavior (\ref{prayleighx}) at small to moderate heights
and the asymptotic exponential behavior (\ref{asympt}) at large heights. However, systematic deviations do exist, which are especially visible at larger values of $\varepsilon$, corresponding to smaller values of the $N$ (degrees of freedom) parameter. These systematic deviations are in large part due to the fact that the true wave height distribution for any given set of input parameters exhibits spatial dependence, evolving from the original Rayleigh distribution imposed by incoming boundary conditions to the broader K-distribution, and then gradually back to a Rayleigh distribution as the wave energy is transferred to longer wavelengths and the steepness decreases~\cite{janssenherbers}. An example of this spatial dependence appears below in figure~\ref{figspatial}. Thus, a more accurate model
for the total wave height distribution consists of a sum of several K-distributions, or equivalently the tail of the
full distribution may be modeled by a K-distribution multiplied by a prefactor $C<1$, as discussed in reference~\cite{yingkaplan}. Nevertheless, as seen in figure~\ref{fignldistr}, equation~(\ref{kbess2}) correctly describes wave height probabilities at the $\pm 20\%$ level of accuracy, allows for an extremely simple one-parameter characterization of the wave height distribution, and facilitates easy comparison between the effects of linear and nonlinear focusing.

\subsection{Scaling with Input Parameters}

\begin{figure}[ht]
\centerline{\includegraphics[width=3.5in,angle=0]{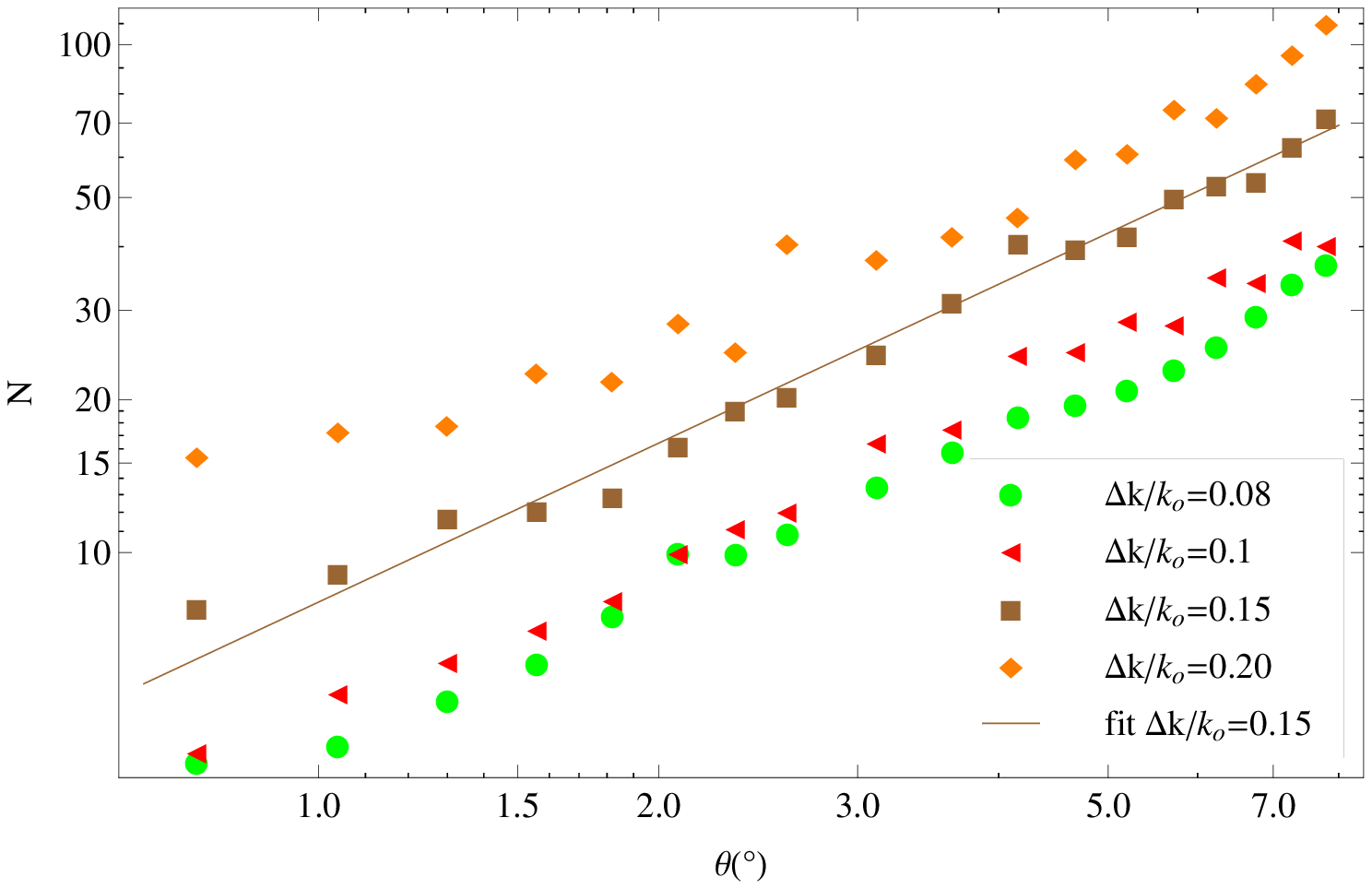}
\includegraphics[width=3.5in,angle=0]{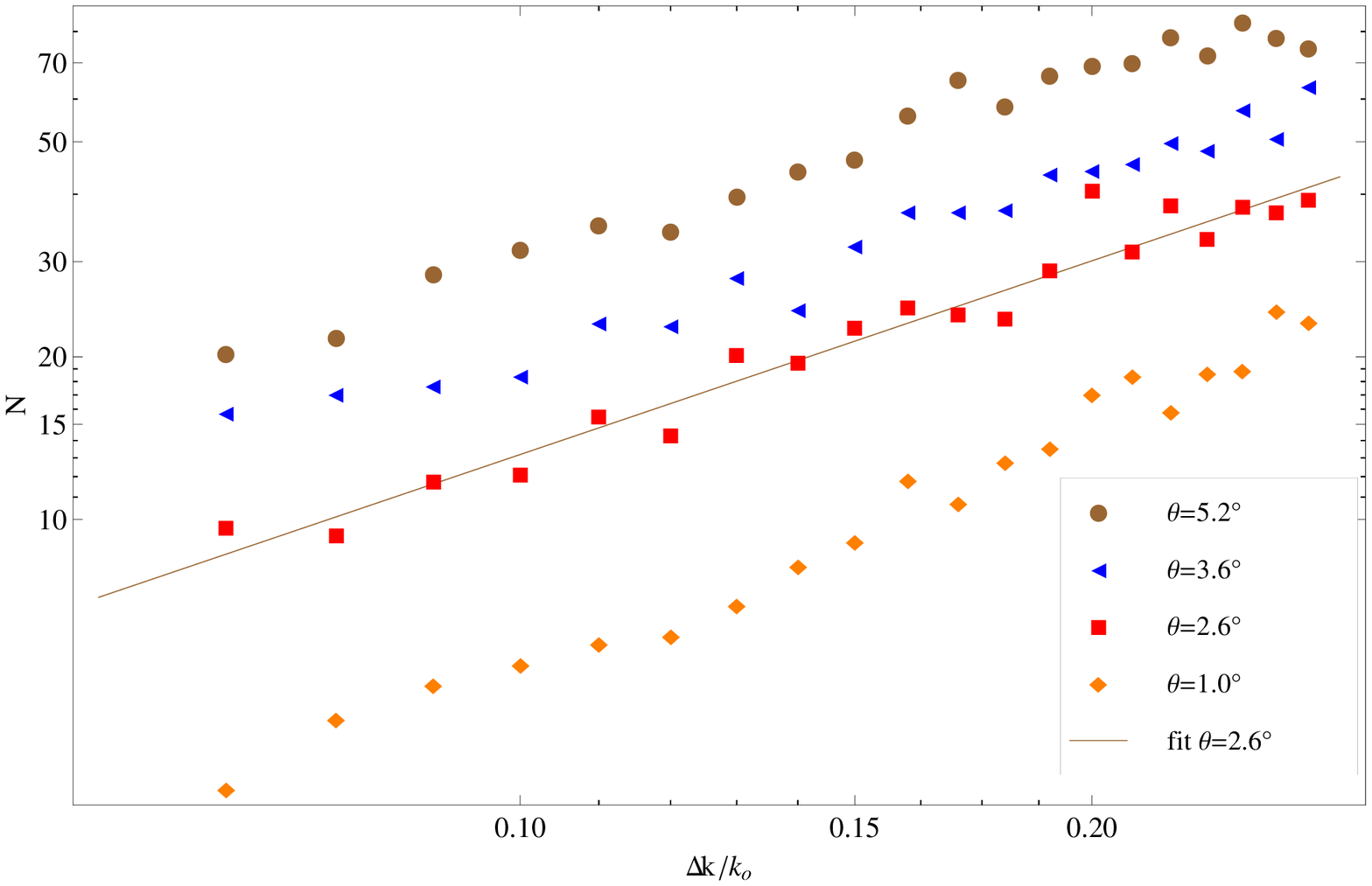}}
%\vskip 0.2in
\caption{The best-fit $N$ value (equation~(\ref{kbess2})) describing the wave height probability distribution is shown as a function of the initial angular spread $\Delta \theta$ and initial wave number spread $\Delta k/k_0$ of the incoming sea. The steepness is fixed at $\varepsilon=0.032$. 
The left panel shows the scaling of $N$ with $\Delta \theta$, with the line showing the best-fit scaling $N\sim (\Delta \theta)^{1.04}$ for $\Delta k/k_0=0.15$. The right panel shows the scaling of $N$ with $\Delta k/k_0$, with the line showing the best-fit scaling $N \sim (\Delta k/k_0)^{1.15}$ for $\Delta \theta=2.6^\circ$.
}
\label{fignlscaling}
\end{figure}

Given the single-parameter approximation of equation~(\ref{kbess2}), it is sufficient to explore the dependence of the parameter $N$ on the input variables describing the incoming sea, specifically the initial angular spread $\Delta \theta$, the initial wave number spread $\Delta k/k_0$, and the initial steepness $\varepsilon$. In the two panels of figure~\ref{fignlscaling}, we fix the steepness at $\varepsilon=0.032$ and show the scaling of $N$ with $\Delta \theta$ and $\delta k/k_0$, respectively. Given that the Benjamin-Feir instability for a monochromatic wave in one dimension~\cite{bf} is at the root of the nonlinear instability in the general case, it is not surprising that stronger deviations from the Rayleigh model, as indicated by smaller values of $N$, occur as $\Delta \theta$ or $\Delta k$ is reduced, consistent with earlier results~\cite{onorato01, dysthe00, socquet05, gramstad07}. Specifically, we find 
\begin{equation}
N \sim (\Delta \theta)^a \left(\frac{\Delta k}{k_0}\right)^b
\label{nonlindtdk}
\end{equation}
where $a$, $b \approx 1$. 

We note that the scaling of $N$ with incoming angular spread $\Delta \theta$ for nonlinear focusing, $N \sim \Delta \theta$, is only half as strong as the scaling $N \sim (\Delta \theta)^2$ arising from linear wave scattering by currents, as implied by equations~(\ref{gammadef}) and (\ref{nval}). Thus, smaller angular spreads $\Delta \theta$ are needed for the nonlinear focusing mechanism to be effective, as compared with linear focusing by currents. 
This is easily seen by comparing the range of $\Delta \theta$ in figure~\ref{fignlscaling} with the corresponding range in table~\ref{enhancement} for the linear mechanism.
On the other hand, figure~\ref{fignlscaling} and equation~(\ref{nonlindtdk}) both imply that the nonlinear mechanism exhibits significant sensitivity to the spectral width $\Delta k/k_0$, consistent with previous findings~\cite{kharif, clamond02, henderson99,lake77,tanaka90,zak06}. This is to be contrasted with the linear mechanism of rogue wave formation, which is insensitive to the spectral width at leading order in $\Delta k/k_0$~\cite{hkd}.

\begin{figure}[ht]
\centerline{\includegraphics[width=4in,angle=0]{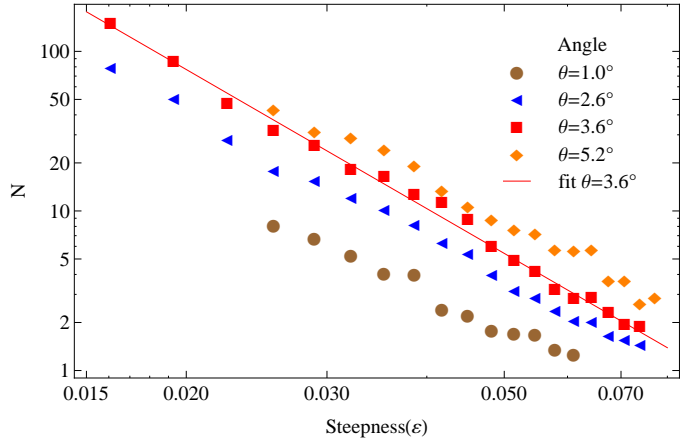}}
%\vskip 0.2in
\caption{The best-fit $N$ value (equation~(\ref{kbess2})) describing the wave height probability distribution is shown as a function of the steepness $\varepsilon$ for several values of the initial angular spread $\Delta \theta$. The
initial wave number spread is fixed at $\Delta k/k_0=0.1$, as in figure~\ref{fignldistr}. The line shows the best-fit scaling $N \sim \varepsilon^{-2.9}$ for $\Delta \theta=2.6^\circ$.
}
\label{fignlsteep}
\end{figure}

Finally, in figure~\ref{fignlsteep}, we fix the steepness $\Delta k/k_0=0.1$, as in figure~\ref{fignldistr}, and examine the scaling of the $N$ value with the steepness $\varepsilon$, for several values of the initial angular spread $\Delta \theta$. As $\varepsilon$ grows, $N$ decreases, indicating greater deviations from the Rayleigh distribution. Again, we observe good power-law scaling with the steepness in the range of parameters considered here. We have 
\begin{equation}
N \sim \varepsilon^c
\label{nonlineps}
\end{equation}
where $c \approx -3$. At larger values of the steepness (not shown), saturation occurs.

\begin{table}[ht]
\begin{center}
%\begin{tabular*}{0.80\textwidth}{@{\extracolsep{\fill}} | c | c | c |}
\begin{tabular}{| c | c | c |}
  \hline
  $\;\;\;\;\;N\;\;\;\;\;$ & \;\;\;\;\;$E(2.2)$\;\;\;\;\; & \;\;\;\;\;$E(3.0)$\;\;\;\;\; \\
  \hline
 2 & $1.1 \cdot 10^2$ & $5.2\cdot 10^4$ \\
  5 & 37 & $7.3\cdot 10^3$ \\
 10 & 16 & $1.3 \cdot 10^3$ \\
 20 & 6.8 & $2.2 \cdot 10^2$ \\
 50 & 2.9 & 27 \\
 100 & 1.8 & 7.8 \\
 \hline
\end{tabular}
\end{center}
\caption{The enhancement in the probability of rogue wave formation (wave height $2H=2.2\, {\rm SWH}$) as well as the enhancement of the probability of extreme rogue wave formation
(wave height $2H=3.0\, {\rm SWH}$) are calculated for several values of the $N$ parameter, as in table~\ref{enhancement}.
}
\label{enhnl}
\end{table}

Table~\ref{enhnl},
calculated analogously to table~\ref{enhancement} in the previous section, aids in extracting the implications
of figures~\ref{fignlscaling} and \ref{fignlsteep} by indicating the quantitative relationship between the $N$ value and the enhancement in rogue wave and extreme rogue wave occurrence probabilities. We note that even at $N$ values between $50$ and $100$, corresponding to the upper range of values in figures~\ref{fignlscaling} and \ref{fignlsteep}, the occurrence of extreme rogue waves is enhanced by an order of magnitude. Exponentially larger enhancement is predicted for parameters associated with smaller values of $N$.

\section{Combined Effect of Nonlinear and Linear Focusing}
\label{seccombined}

Finally, we discuss the possibility of even greater enhancement in the rogue wave formation probability when linear and nonlinear mechanisms are acting together~\cite{janssenherbers,yingkaplan}. In this context, it is important to consider again the spatial scales associated with rogue wave development in the two mechanisms. We recall that when an incoming random sea is linearly scattered by strong currents, the first singularities in the ray dynamics occur after a distance scale $L \sim \xi (u_{\rm rms}/v)^{-2/3}$, as discussed in section~\ref{secrayinten}. These first singularities are the ones associated with the highest probability of rogue wave formation, as subsequent random scattering exponentially stretches the phase space manifold and leads to ever smaller density associated with each second- and higher-order singularity~\cite{lkbranch,hkd}. At distances $\gg L$, the pattern of hot and cold spots becomes less and less prominent, the ray density again becomes nearly uniform (see figure~\ref{figrayimage}), and the wave height distribution asymptotically approaches again the Rayleigh limit.
Similarly, nonlinear evolution as described by the NLSE without current (equation~(\ref{cnls4}) with $\Phi_c=0$) occurs on a typical distance scale $1/k\varepsilon^2$. On distance scales larger than $1/k\varepsilon^2$, energy transfer from smaller to larger wavelengths (i.e., the frequency downshifting effect mentioned previously in Sec.~\ref{secnlse}) results eventually in a decline in the steepness and again an approach towards the limiting Rayleigh distribution~\cite{janssen09,tanaka01,gibson07}.

\begin{figure}[ht]
\centerline{\includegraphics[width=4in,angle=0]{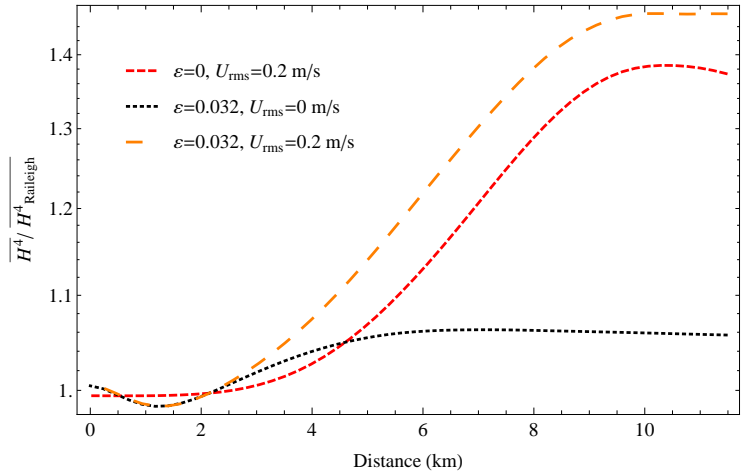}}
%\vskip 0.2in
\caption{The fourth moment of the wave height distribution is shown as a function of evolution distance, starting in each case from a Longuet-Higgins random sea with mean
wave speed $v=7.81\,{\rm m/s}$, initial angular spread $\Delta \theta=5.2^\circ$, and wave number spread $\Delta k/k_0=0.1$. The three situations considered are:
(a) linear scattering by random currents of rms speed $u_{\rm rms}=0.2$~m/s and eddy correlation length $\xi=800$~m,
(b) nonlinear evolution with initial steepness $\varepsilon=0.032$ and without currents,
 and (c) nonlinear evolution in the presence of currents.
}
\label{figspatial}
\end{figure}

This behavior is illustrated in figure~\ref{figspatial} for linear evolution with random currents, nonlinear evolution in the absence of currents, and for a scenario in which the two mechanisms are both active. Here we use the fourth moment $\overline{H^4}$ as a convenient measure of the size of the tail of the wave height distribution. Note that for the chosen parameters, the distance scales associated with linear and nonlinear rogue wave formation are comparable. Clearly, in this case currents have a greater effect than nonlinear focusing, but the strongest deviations from Rayleigh statistics are observed when linear scattering and nonlinear interaction are both present.

\begin{figure}[ht]\centerline{\includegraphics[width=3.0in,angle=0]{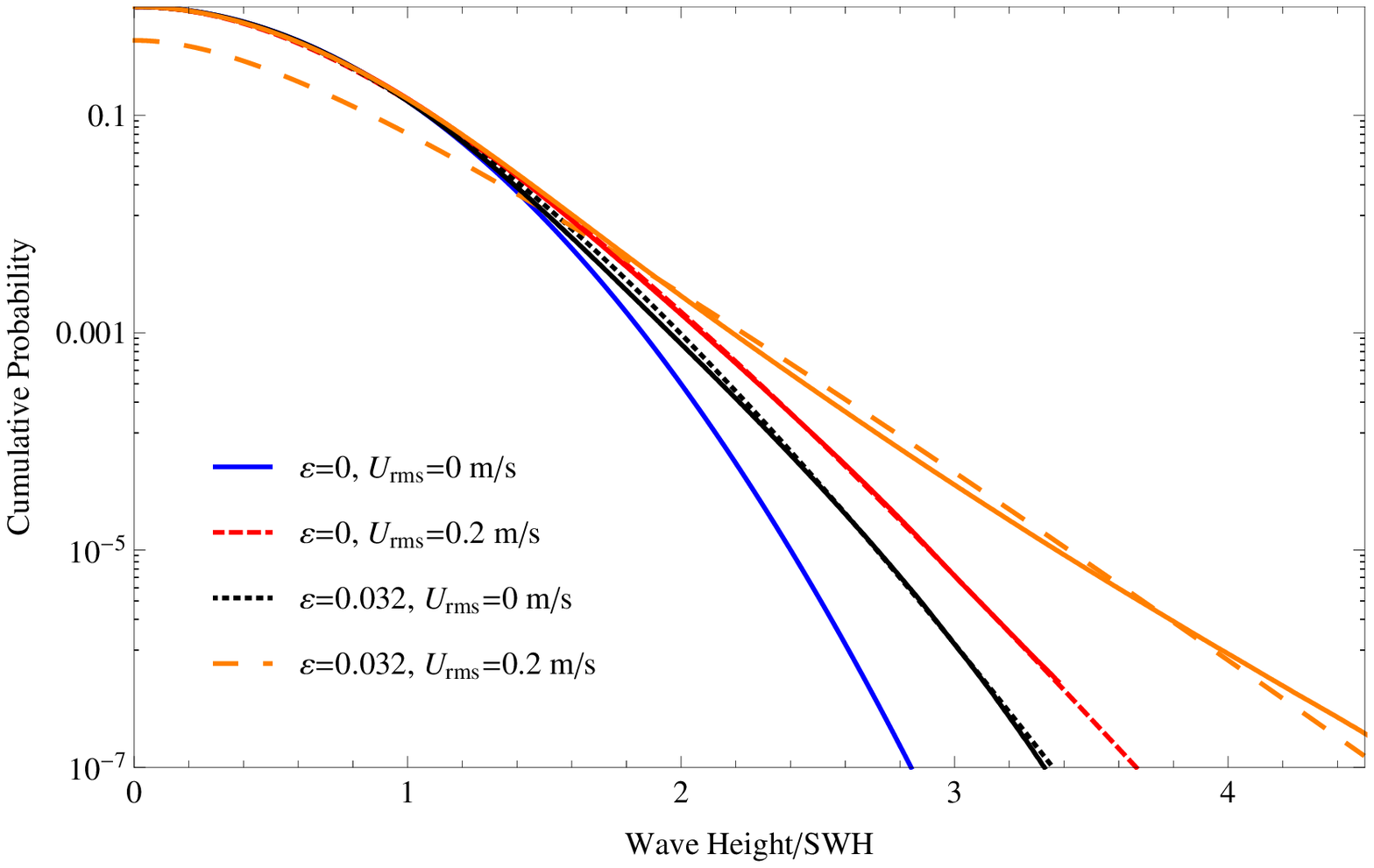} \includegraphics[width=3.0in,angle=0]{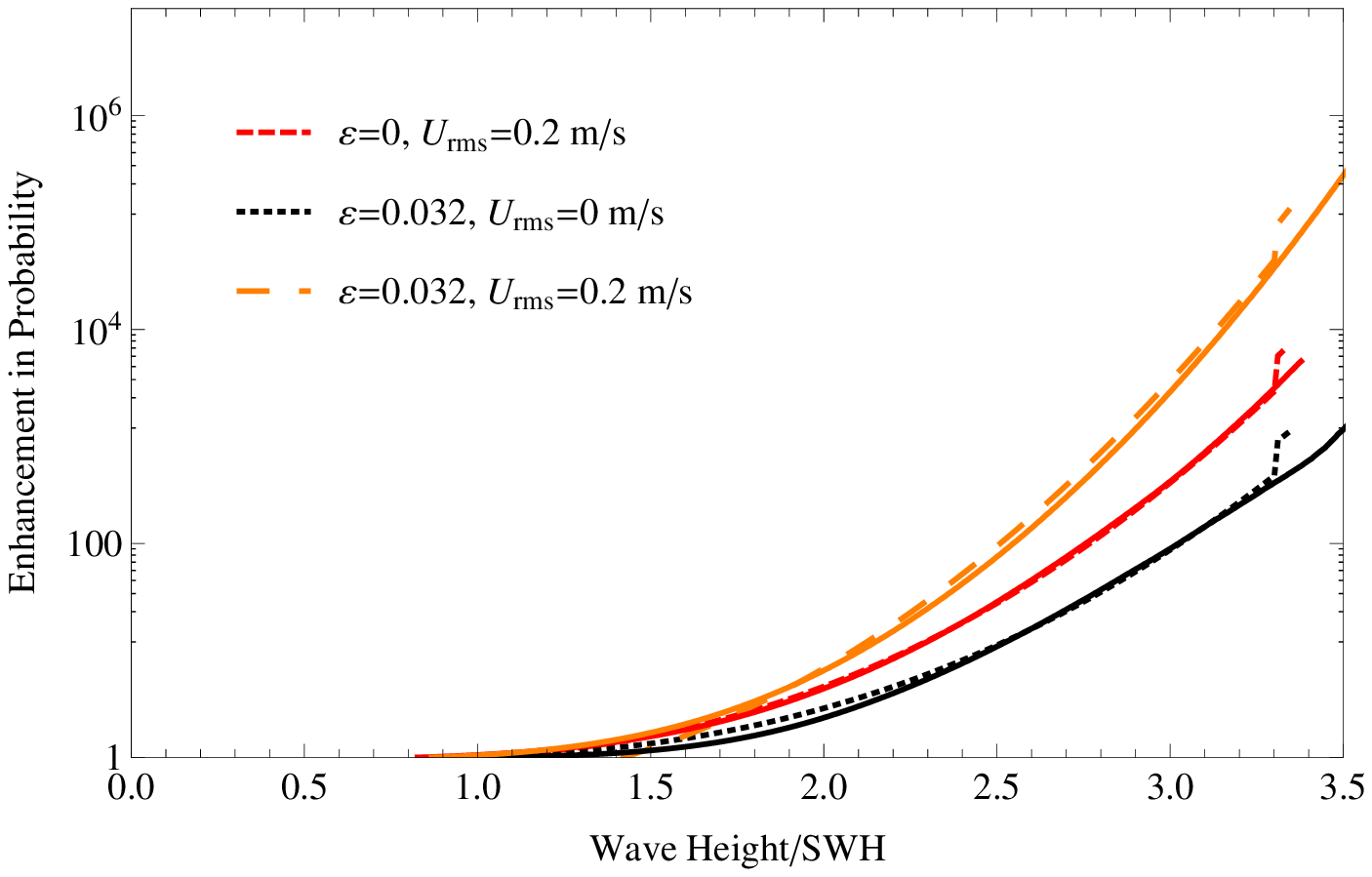}}
%\vskip 0.2in
\caption{Left panel: The cumulative distribution of wave heights, in units of the significant wave height, is obtained for the same three scenarios as are considered in figure~\ref{figspatial}. In each case, the solid curve is a fit to a K-distribution with (a) $N=16$ for linear scattering by currents, (b) $N=29$ for nonlinear evolution, and (c) $N=5.1$ when linear and nonlinear focusing are acting in concert. The Rayleigh distribution ($N= \infty$) is shown for reference. Right panel: In each of the three scenarios, the probability enhancement factor $P_{\rm total}(H)/P_{\rm Rayleigh}(H)$ is obtained from the data.
}
\label{fignonlincurrent}
\end{figure}

The total wave height distributions for these same three scenarios, and the probability enhancement over the predictions of the Longuet-Higgins model, are shown in figure~\ref{fignonlincurrent}. As noted above in section~\ref{secnlheight}, when wave height data is collected over a large spatial field that includes some areas of very strong deviations of Rayleigh statistics and other areas where such deviations have not yet had an opportunity to develop, the full distribution may not be well approximated by a single K-distribution, but the tail may still be well approximated in this way, since it is dominated by data from those areas where deviations are strongest~\cite{yingkaplan}. This is indeed what we clearly observe in figure~\ref{fignonlincurrent}, for the scenario where nonlinearity and currents are both present.

Again we see from figure~\ref{fignonlincurrent} that deviations from Rayleigh statistics become ever more pronounced as taller and taller waves are considered, as expected from the asymptotic form of the K-distribution (equation~(\ref{asympt})). In particular, in this example we see that the probability of forming an extreme rogue wave (wave height $=3$~SWH) is enhanced by a factor $90$ due to nonlinear interaction, by a factor of $380$ due to focusing by currents, and by a factor of $2600$ when the two mechanisms are combined.

\section{Conclusions and Outlook}

It will take some time to sort out the mechanisms of rogue wave formation with complete certainty. All potentially important factors and mechanisms ought to be included in the discourse, which we hope will someday lead to agreement about the several formation mechanisms and their interactions. More importantly, predictive tools leading to safer navigation should eventually emerge. One of the seemingly important factors, which might be called ``statistical focusing,'' is highlighted here. In terms of wave propagation, statistical focusing is a linear effect (although it is nonlinear dynamics at the level of ray tracing). It leads to large enhancements in the frequency of rogue wave formation under reasonable sea state assumptions.

Statistical focusing combines the effects of deterministic wave refraction by current eddies with Longuet-Higgins statistical ideas under realistic conditions. The key notion is that the focusing effects of eddies, which would be very dramatic on an (unrealistic) monochromatic and unidirectional sea, are not altogether washed out when realistic frequency and directional dispersion are included. Essentially, deterministic caustics present in the unrealistic idealization are smoothed into hot spots, which are then treated statistically within Longuet-Higgins theory. The hot spots dominate the statistics in the tail of the wave height distribution. This amounts to a nonuniform sampling version of Longuet-Higgins theory, with a solid basis for the nonuniform energy density distributions used.

Since nonlinear effects are also important, we have examined them alone within the popular fourth-order nonlinear Schr\"odinger equation (NLSE) approximation for nonlinear wave evolution under realistic seaway conditions. Finally, we have investigated the combined effect of nonlinear wave evolution and statistical focusing. We find that strongest deviations from Rayleigh statistics are observed when linear scattering (statistical focusing) and nonlinear interaction (NLSE) are both present. However, for the parameters chosen here at least, the linear scattering due to eddies was more important than the nonlinear effects, which require large steepness or a very narrow range of propagation directions to become significant.

We have presented a measure closely related to the probability of rogue wave formation, the freak index $\gamma$. This could conceivably become the basis for a probabilistic forecast of rogue wave formation, in the spirit of rainfall forecasts.

There are at least three clear directions for future development of the work presented here. First, both the computer simulations and the theory must be developed further to explore fully and systematically the combined effects of nonlinear and linear focusing. This will also involve investigating in depth the underlying mechanism through which the formation of hot and cold spots is aided by nonlinear focusing. Secondly, a better understanding is needed of the stability of the hot spot patterns under slow changes in the current field or in the spectrum or directionality of the incoming sea. The strength of what might be called scintillation or twinkling~\cite{twinkling} in analogy with the case of light traveling through the atmosphere will have important consequences for the predictive power of the model. Thirdly, and most importantly, there is a clear need to compare the model simulations with observations and experiments. Although comprehensive global data are not available at this point, it may be possible to compare the results to local observations where data are more readily available, e.g., in the North Sea.

Whatever the final word is on rogue wave formation (or final words, because there may be more than one mechanism), it must involve a reallocation of energy from a larger area to a smaller one. Waves cannot propagate and increase in height at no expense to their neighbors: the energy has to come from somewhere, and the effect must be to reduce the wave energy somewhere else. The focusing mechanism is clear in this respect: hot spots form and cold spots do too, according to a ray tracing analysis, maintaining energy balance~\cite{brethertongarrett}.

\ack This work was supported in part by the US
NSF under Grant PHY-0545390.

\end{document}